# Workshop on Aesthetics of Connectivity for Empowerment at ACM Designing Interactive Systems 2024

Jun Hu[1, *], Mengru Xue[2], Cheng Yao[2], Yuan Feng[3], Jiabao Li[4], Preben Hansen[5]

1. Eindhoven University of Technology, Eindhoven, The Netherlands
2. Zhejiang University Ningbo Innovation Center, China
3. Northwestern Polytechnical University, China
4. University of Texas at Austin, United States
5. Stockholm University, Sweden

Connectivity enabled by technologies such as the Internet of Things, Artificial Intelligence, Big Data, and Cloud Computing is rapidly transforming our interactions with the world and with each other. It reshapes social interactions, fostering collaboration, creativity, and unprecedented access to information and resources. However, this connected world and era demand innovative design approaches that harmonize technical functionality with human-centered values. We have run a series of workshops at different conferences, trying to engage the participants in discussions about the related challenges and opportunities, of digital art [1] and aesthetics [2] to AI-driven creativity [3] and their functional aspects in healthcare [1] and empowerment [2, 3]. We want to focus further on the intersection of these challenges where we see opportunities: leveraging *aesthetics* and *connectivity* as catalysts for *empowerment*.

We organized a workshop [4] on this topic at the DIS 2024 (ACM Designing Interactive Systems 2024) conference. This workshop explores how design can look into not only the pragmatic quality of connectivity but also the aesthetic aspects and their functional qualities for empowering individuals or groups through sensory engagement, emotional resonance, and social connectedness. Aesthetics, not only a subject in the science of sensual perception but also a mechanism for designing intelligent systems [5-7], can serve as a link between humans and technology to improve cognitive function, reduce stress, and inspire creativity, while connectivity integrates physical and digital realms to address complex human values and social needs. When combined, these elements empower individuals to reimagine, explore, and redefine their interactions with technology, environments, and each other.

The nine papers presented in this workshop tried to touch upon the topics from different perspectives across different application areas, demonstrating the transformative potential of integrating aesthetics and connectivity and utilizing aesthetics enabled by connectivity for empowerment.

## Connectivity and supporting technologies

The papers explore a variety of connectivity-enabling technologies empowerment. Mulundule et al [8] used Smart clothing with embedded sensors and conductive materials for its ability to monitor and respond to physiological data. Magicarpet [9] features technologies such as interactive play carpets equipped with sensor switches and LED panels, whereas Figame [10] is developed using the Unity engine, which incorporates Joint Media Engagement (JME) strategies to facilitate healthy gaming relationships between parents and children. Emotional well-being and stress management are addressed through various tools. These tools are, for example, deepfake voice technology in InnerSelf [11], wearable stress measurement devices integrated with VR spaces for colleague students [12], and robotic interactions using embedded sensors and machine learning algorithms to detect and manage stress in Children with the Petting Pen [13]. In the other two papers, to enable client-driven decision-making and align the design outcomes with user needs and preferences, AI-generated content (AIGC) and empathy-driven design tools are used for co-design and user engagement [14, 15]. These diverse technologies collectively contribute to enhancing connectivity and empowerment across various contexts.

## Aesthetic experience explored

The papers explore a variety of aesthetic experiences aimed at enhancing emotional and sensory engagement. In the process of co-design and user engagement in designing an authentic Airbnb experience, the focus is on authenticity and bespoke design processes, emphasizing the importance of clients expressing their unique values and preferences [14]. Smart clothing [8] combines aesthetics with technological functionality, by taking into consideration both the visual appeal and the user needs and preferences. Interactive play carpets [9] and digital games [10] are designed to engage parents and children in healthy gaming relationships. Although the voice is deepfacked in InnerSelf [11], since people are familiar with their own voice, the familiarity makes the experience more comforting and therefore, appealing. The calming visuals, soothing sounds, and immersive environments in VR spaces play a crucial role in stress reduction [12]. The interaction with the Petting Pen, including aesthetic aspects in its appearance, movements, and the way it communicates, can influence the child's emotional response [13]. Sensory-rich experiences are created in immersive virtual nature environments for relaxation and recovery [16]. Also, the integration of traditional calligraphy with real-time physiological data in mindfulness-training tools enhances emotional well-being through interactive and multimodal experiences [15]. These diverse aesthetic experiences are influenced not only by multimodal sensual contact with the connectivity but also by the feelings and values in interacting with it.

## Empowerment

Empowerment is explored in several different ways in the papers through the integration of aesthetics and connectivity. To engage the users in co-design, An's work tried to enable them to express their unique values and preferences, thereby fostering authenticity and creativity in design processes [14]. Ma et al. try to empower users by enhancing their physical and emotional well-being through combining aesthetics with the functionality of smart fabrics [8]. Values of social connectivity and engagement in interactive play and communication, particularly for children with autism and their parents, are the focus areas of interactive platforms and games [9, 10]. Users are empowered with personalized assistance and emotional regulation through deepfake voice technology and mindfulness training systems [11, 15], where personalization improves the aesthetic experience by making it more meaningful and relevant to the individual. Relaxing and restorative experiences in immersive virtual environments help users manage stress for their mental health [16]. Children's awareness and coping skills are enhanced with stress measurement and management tools for their overall well-being [13]. Overall, these efforts contribute to empowering individuals by enhancing their emotional, social, and physical well-being through the thoughtful integration of aesthetics and connectivity.

## Considerations and Challenges

From these papers, we observe various considerations and challenges in applying the aesthetics of connectivity for empowerment. In An's work of engaging users in co-designing Airbnb experiences [14], the primary considerations include ensuring authenticity and avoiding consumerist aesthetics, which can lead to a loss of personal values and cultural authenticity. It is, however, challenging to balance or, if possible, combine the aesthetic preferences of the designers with the unique values of the clients, overcoming the possible biases introduced by AI-generated content.

Mulundule et al. need to consider how to integrate aesthetics with technological functionality, ensuring user comfort and addressing environmental adaptability while facing the challenges in managing the complexity of smart clothing design, production issues, and ensuring cultural sensitivity and inclusivity [8].

In designing Magicarpet [9], Hu et al.'s considerations are more on fostering parent-child interaction and engagement, particularly for children with autism, where the challenge is to design an inclusive game for children with different levels of abilities and ensure timely feedback and motivation.

When designing an immersive virtual nature, selecting and utilizing natural elements to promote relaxation and recovery need to be considered. Challenges are mostly related to "nature". It includes ensuring the effectiveness of virtual nature scenes in stress recovery and addressing individual differences in user perception of these scenes [16].

The emotional well-being and stress management applications need to take into account personalized and supportive self-talk experiences and the effectiveness of mindfulness-training tools [11, 15]. Challenges are mostly technical, involving accurately detecting and responding to users' emotional states and integrating multimodal feedback systems [11, 15]. When designing stress management tools for children, Li et al. pay more attention to unobtrusive stress detection and relaxing interactions [13]. Accurately capturing stress-related behaviors and integrating behavioral and bio-sensing capabilities are similar challenges.

These considerations and challenges highlight the complexities and opportunities in using aesthetics of connectivity for empowerment across different application areas.

# Future research

Future studies can further investigate a number of interesting research directions described in these papers. It is well known already that personalized and adaptive systems, such as context-aware wearables and emotionally intelligent interfaces, can enhance the user experience by adapting to individual needs in real-time. Inclusive and accessible principles can be used in designing interactive products and systems for users with different abilities and cultural backgrounds. Multisensory input and biofeedback mechanisms can help to improve user engagement and emotional well-being. Co-design and user participation methods can ensure that user values and preferences are considered sufficiently in the design process, fostering empathy-driven design.

When we called for papers with the word "connectivity" in the title, we expected the submissions to address more the distributed systems that would involve interconnected products and systems, and further people and their societies. However, almost none of the papers addressed such a topic. There are several aspects were not well-covered in these papers. None of the papers undertook longitudinal studies that are necessary to evaluate the long-term impact of connectivity and related aesthetics and technologies on behavior, emotional health, and social dynamics. Most of these papers did not discuss ethical and sustainable issues, such as environmentally friendly materials and ethical AI practices. These issues should be addressed more in the design and research processes.

These research directions will contribute to designing more inclusive, adaptive, and user-centered products, systems, and services by addressing the evolving needs and challenges in the fields of connectivity, aesthetics, and empowerment.